\documentclass[conference]{IEEEtran}
\usepackage{cite}
\usepackage{amsmath,amssymb,amsfonts}
\usepackage{algorithmic}
\usepackage{graphicx}
\usepackage{textcomp}
\usepackage{xcolor}
\usepackage{url}
\usepackage{float}

\ifCLASSOPTIONcompsoc
    \usepackage[caption=false,font=normalsize,labelfont=sf,textfont=sf]{subfig}
\else
    \usepackage[caption=false,font=footnotesize]{subfig}
\fi

\def\BibTeX{{\rm B\kern-.05em{\sc i\kern-.025em b}\kern-.08em
\\    T\kern-.1667em\lower.7ex\hbox{E}\kern-.125emX}}

\title{A deep learning based tool for automatic brain extraction from functional magnetic resonance images in rodents}

\author{\IEEEauthorblockN{Sidney Pontes-Filho\IEEEauthorrefmark{1}\textsuperscript{,}\IEEEauthorrefmark{2}, Annelene Gulden Dahl\IEEEauthorrefmark{3}, Stefano Nichele\IEEEauthorrefmark{1} and Gustavo Borges Moreno e Mello\IEEEauthorrefmark{1}\textsuperscript{,}\IEEEauthorrefmark{4}}
\IEEEauthorblockA{\IEEEauthorrefmark{1}\textit{Department of Computer Science, Oslo Metropolitan University, Oslo, Norway}\\
\IEEEauthorrefmark{2}\textit{Department of Computer Science, Norwegian University of Science and Technology, Trondheim, Norway}\\
\IEEEauthorrefmark{3}\textit{Kavli Institute for Systems Neuroscience, Norwegian University of Science and Technology, Trondheim, Norway}\\
\IEEEauthorrefmark{4}\textit{Department of Mech., Elec. and Chem. Engineering, Oslo Metropolitan University, Oslo, Norway}\\
Email: gustavom@oslomet.no}
}

\begin{document}
\maketitle
\begin{abstract}

Removing skull artifacts from functional magnetic images (fMRI) is a well understood and frequently encountered problem. Because the fMRI field has grown mostly due to human studies, many new tools were developed to handle human data. Nonetheless, these tools are not equally useful to handle the data derived from animal studies, especially from rodents. This represents a major problem to the field because rodent studies generate larger datasets from larger populations, which implies that preprocessing these images manually to remove the skull becomes a bottleneck in the data analysis pipeline. In this study, we address this problem by implementing a neural network based method that uses a U-Net architecture to segment the brain area into a mask and removing the skull and other tissues from the image. We demonstrate several strategies to speed up the process of generating the training dataset using watershedding and several strategies for data augmentation that allowed to train faster the U-Net to perform the segmentation. Finally, we deployed the trained network freely available.

\end{abstract}


\section{Introduction}

Functional magnetic imaging (fMRI) has emerged as a powerful tool to investigate functional networks in the brain. Because fMRI is a non-invasive technology, the field has primarily been driven by its application to the study of the human brain. Consequently, great advances in automating analysis of fMRI data through tools that improve its speed and efficiency have been achieved to process human data, saving both time and costs associated with fMRI studies. However, efforts to either modify preexisting tools, or develop similar tools for use on rodent datasets are lagging. 

Currently, one of the most time consuming steps in the processing of rodent fMRI data is the process of brain extraction or skull stripping. This step consists of segmenting the whole brain, which is equivalent to removing all non-cerebral tissue, including the skull, nose, mouth, ears, and muscles \cite{hahn2000skull}. Accurate extraction of the brain is essential to ensure that fMRI data of all the subjects in the study are anatomically aligned, which is necessary to allow for reliable statistical comparison across large cohorts of animals \cite{spring2007sexual,bearer2009reward,vousden2015whole,zhang2010altered,van2013mapping}. Because skull stripping is a well understood problem \cite{kapur1996segmentation} and a necessity in every fMRI analysis, the development of tools to automatise and increase the speed and reliability of results might have a great positive impact into fMRI research.

Rodent's brain extraction poses additional challenges when compared to segmenting the human brains from fMRI data. Rodents have a smaller gap between the brain and the skull, resulting in a less clear edge demarcation than in humans. Additionally, the rodent brain differs in shape, texture, size and proportion from the human brain. This means that the automated tools developed to handle human data such as Brain Extraction Toolkit (BET) \cite{smith2002fast} and BrainSuite’s Brain Surface Extractor (BSE) \cite{shattuck2002brainsuite} usually fail to process images of rodent brains. Therefore, brain extraction of rodent anatomical and functional data is predominantly carried out manually. This process involves researchers going slice-by-slice through the acquired (anatomical and functional) images in all three dimensions and manually drawing masks for the brain using a mouse or a tablet. 

A tool to efficiently extract the brain from rodent anatomical images was recently published\cite{delora2016simple}. This tool takes as input one representative brain from the study and its manually created brain mask, and uses this information to carry out the brain extraction of the remaining subjects in the study. While this is a great tool for extracting the brains in the anatomical images, it is not intended for use in functional datasets, and there is, to the best of our knowledge, no equivalent tool available for extracting the brain from the functional dataset. In order to observe the changing activity of the brain over time, the functional datasets have to be acquired at a much greater speed than the anatomical images, resulting in a much lower spatial resolution than the anatomical images. To preserve the sensitivity to blood-oxygenation-level-dependent (BOLD) contrast the images are also frequently subject to severe susceptibility-induced distortions, in particular, in the back of the brain near ear canals and sinuses. Due to these confounds, skull extraction of functional rodent images commonly fails, and the current state-of-the-art in the field of rodent imaging is to manually draw the masks. This process is both time-consuming and often inaccurate, contributing to a less-than-perfect alignment of the functional data to the template brain.

To overcome this obstacle, we have developed a deep learning-based tool in Python that quickly and successfully extracts the brain from the functional datasets, thus improving the speed and accuracy of the preprocessing pipeline. The tool, furthermore, does not require any study-specific input from the researcher in order to successfully separate brain from non-brain tissue.  The tool is freely available online.
%To accommodate or facilitate the use of the tool to studies that use different scanners, sequences and different acquisition parameters; we have added a graphical user interface that can be modified to each study, thus improving the study-specific accuracy and ease of use. We have verified the tool on functional images from not only our lab but also another lab using a different pre-clinical scanner, pulse sequence and acquisition parameters.

\section{Related works}
In the last three decades, many methods for skull stripping have been proposed \cite{eskildsen2012beast,galdames2012accurate,speier2011robust}, ranging from simple luminance thresholding to 3D-convolutional deep learning techniques\cite{hwang20193d}. Among them, the most promising are the water-shedding based segmentation\cite{hahn2000skull}, the Brain Extraction Tool (BET)\cite{smith2002fast} and the most recent 3D-U-Net\cite{hwang20193d}. Watershed based methods are image processing pipelines originally described in \cite{hahn1996vegetation} that are advantageous for being unsupervised, fast, and easy to tune; they leverage luminance gradients to define regions of interest that can be defined either as brain or non-brain. BET, on the other hand, uses a malleable model, where a spherical mesh is initialized at the center of mass and then expanded towards the surface of the brain; locally adaptive model forces based on local intensity values guide this process, allowing BET to quickly segment the brain. The caveat is that BET has a spherical (human) brain assumption, and has irregular performance with oblong elliptical shaped brains, such as rodent brains. Finally, 3D-U-Net is a promising robust methodology that uses convolutional neural networks to perform semantic binary segmentation. This method has the advantage of being able to learn from experts by mapping spacial features of the raw fMRI image to ground-truth data generated by manual segmentation. Because of the need for coregistration and alignment in the z-axis, this method cannot benefit from several of the data augmentation methods available, such as elastic transformations \cite{moshfeghi1991elastic,ronneberger2015u}, thus requiring much more data than the standard U-Net \cite{ronneberger2015u}. All of these methods were developed to handle human fMRI data, and regardless of the great levels of performance achieved by the previously cited methods, a solution to reliably perform skull stripping in rodent data is still missing.

The solutions to particularly handle rodent fMRI data use more modest technologies. More often than not, skull stripping is still done by creating hand-drawn masks and only occasionally helped by semi-automation tools such as BrainSuite's Brain Surface Extractor (BSE)\cite{shattuck2002brainsuite} which produces an initial mask that subsequently needs to be refined and corrected by hand. Beyond BSE other two automation method categories are available, warping to brain atlas based methods, and surface template based methods \cite{delora2016simple}. Both methods are built extending the NiftyReg software package \cite{modat2010lung}; and both dependent on the warping of the image to a template coordinate map, or on warping a mask to the raw image through a series of affine and non-linear transformations. These methods produce excellent results on high-resolution anatomical images, but due to the lower spatial resolution and image distortions in the functional datasets the automated skull stripping methods currently available  fail to perform correctly on rodent functional images. Hence, the brain extraction problem in functional images from rodent data has yet to be solved satisfactorily in a generic and robust way.

%Nonetheless, functional images have a much lower spatial resolution to optimize for temporal resolution. Additionally, functional imaging has often several spatial and temporal artifacts that manifest themselves as image distortions. Because of these factors, and to the best of our knowledge, all available automated skull stripping methods fail to perform correctly on functional images derived from rodents. Hence, the brain extraction problem in functional images from rodent data has yet to be solved satisfactorily in a generic and robust way.

\section{Methods}

\subsection{Image acquisition}
62 fMRI datasets from 31 McGill-R-thy-App rats were acquired on a 7T Biospec 70/30 (Bruker BioSpin) preclinical scanner, equipped with an actively shielded 660 mT/m BGA12S HP gradient set (Bruker) in combination with a quadrature surface coil (Bruker BioSpin). Aspin-echo EPI sequence was used with the following parameters: 600 repetitions (total scan time of 30 min each) with 2 segments, TE= 20ms, repetition time (TR) = 1.5s for a full-volume acquisition of 3s., field-of-view (FOV) of 20x20mm, matrix size 80x80, 55 dummy scans, flip angle of 90 degrees. Seventeen slices were acquired in rostro-caudal direction for a final resolution of 250 x 250 x 1000um. All procedures were approved by the Norwegian Food Safety Authority as well as the local Animal Welfare Committee of the Norwegian University of Science and Technology (NTNU). All animals were housed and handled according to the Norwegian laws and regulations concerning animal welfare and animal research. Experimental protocols were approved by the Norwegian Animal Research Authority (FOTS application number 11932) and were in accordance with the European Convention for the Protection of Vertebrate Animals used for Experimental and Other Scientific Purposes.

\subsection{Training dataset and Watershedding-based brain segmentation}
Due to the success of the watershedding algorithm to segment the brain in human fMRI dataset \cite{malpica1997applying}, we used it as a semi-automated approach to generate a dataset of masks that were used to train the neural network to segment the brain from the skull. 
Watershedding is a region-based approach that considers the target structure as a homogeneous region which is determined by a search process guided by appropriate criteria for homogeneity. We implemented the watershedding segmentation by using functions in OpenCV\cite{opencv_library} to preprocess the images by gray-scaling, mean-shifting and normalizing them. Once the images were considered suitable for segmentation, we thresholded the gray-scaled image into masks, calculated their basin gradients, filtered these gradients, and identified the segmented areas as connected components. As result, the watershed method provides per each image a series of masks for each structure in each image.
The gradient of an image function f is the vector constituted by the partial derivatives in each image dimension. The gradient's direction is the direction of steepest descent and a magnitude ($mag$) is the length of the gradient vector. For an image function in R2, $f(x,y)$ the magnitude of the gradient is calculated as% \eqref{eq0}. 
\begin{equation}
\label{eq0}
%mag(∇f)=√(δfδx)2+(δfδy)2(4.7)
mag(\bigtriangledown f)=\sqrt{(\frac{\delta f}{\delta x})^2+(\frac{\delta f}{\delta y})^2}.
\end{equation}

To choose the mask that represented the brain structure, we leveraged the regularities in the data. Because in this dataset the brain was always very close to the center, this meant that the average polar radial distance between each point of the brain structure mask and the center of the image was shorter than any other structure. Thus, we used this as the criterion to exclude other structures.
The parameters for this process were chosen manually and the results followed by close supervised eye-inspection. Nonetheless, this semi-supervised approach proved to be substantially faster than the manual alternative, because the same parameters could be used for different datasets acquired in similar conditions.

\subsection{Deep-Learning-based segmentation}
In this article, we use a standard U-Net \cite{ronneberger2015u} architecture to perform skull stripping from fMRI images of rodents. U-Nets are most often used for semantic segmentation tasks. Beyond performing well on the task, they allow for efficient use of GPU memory, which is an asset for processing big image datasets with many features. This is heavily dependent on the fully convolutional architecture of the U-Net, which enables the extraction of image features at multiple image scales. In the U-Net, different layers capture coarse feature-maps that reflect this contextual information about the category and location of objects at multiple scales. These feature-maps are later merged through skip connections to combine coarse- and fine-level dense predictions\cite{ronneberger2015u}.

\begin{figure}
  \includegraphics[width=\linewidth]{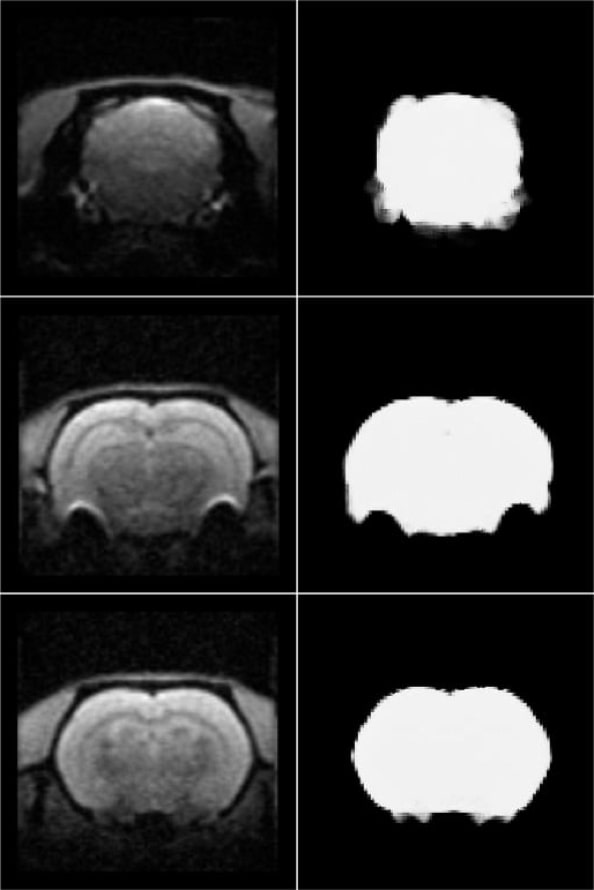}
  \caption{Soft masks examples: The left column represents the input image. The right column illustrates the mask prediction for three different coronal slices of a rodent's  .}
  \label{fig:mask}
\end{figure}

The goal of the U-Net neural network architecture is to predict which pixels in the image matrix are to be classified as brain and which ones are to be classified otherwise. Thus, the output of the final decoder layer is a soft mask (see Fig. \ref{fig:mask}) that when multiplied to the input image produces the final segmented brain region. 

\subsection{Data Augmentation and Training}
One major advantage of using the U-Net is that it is possible and simple to use several methods for data augmentation such as resizing, flipping, rotating, and minor translations. These data augmentation strategies increase the performance of the model by increasing the size and variety of the dataset\cite{simard2003best}. 

Additionally, fMRI images often do have distortions and movement artifacts. To improve U-Net's robustness in face of such artifacts, elastic affine transformations were applied equally to the input image and the target masks. In total, the training and validation datasets were increased by 50\% with these slightly deformed images\cite{simard2003best}.

To speed up training we utilized a U-Net pre-trained to segment pathological structures in human MRI images. Instead of stochastic gradient descent, we modified the original optimizer of U-Net to Adam \cite{kingma2014adam} with the learning rate of 0.001. Additionally, the training used batches of 25 images during 1,000 epochs.

\section{Results}
This section presents the quantitative and qualitative results of the deep learning neural network for our task of segmenting rodents' brains. Table~\ref{table:results} contains the quantitative results of binary cross-entropy (BCE) loss, accuracy, F1 score, precision and recall on the validation dataset of 49 images (5~\% of our dataset). All these values show that our model segments almost all pixels that contain the brain (98.3~\% recall) with precision of 98.5~\%. The F1 score is a metric that combines recall and precision. The accuracy represents the percentage of correct answers for the pixels predicted as part of the brain or not. Such value is high and it is 99.35~\%. Those measurements suggest that the model performance is excellent.

\begin{table}[ht]
\renewcommand{\arraystretch}{1.1}
\centering
\caption{Validation results of the best (lowest) loss.}
\label{table:results}
\begin{tabular}{|l|l|}
\hline
\textbf{Measurement} & \textbf{Value} \\ \hline
BCE loss & 0.01562267541885376 \\ \hline
Accuracy & 0.9935703277587891 \\ \hline
F1 Score & 0.9843953251838684 \\ \hline
Precision & 0.9854521751403809 \\ \hline
Recall & 0.9833407998085022 \\ \hline
\end{tabular}%
\end{table}

\begin{figure*}[ht]
\centering
\subfloat[]{\label{fig:s01}\includegraphics[width=0.30\textwidth]{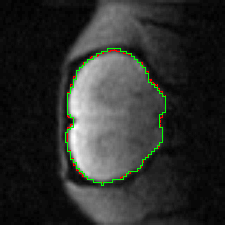}}\hfill
\subfloat[]{\label{fig:s02}\includegraphics[width=0.30\textwidth]{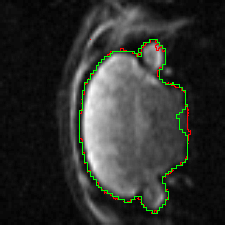}}\hfill
\subfloat[]{\label{fig:s03}\includegraphics[width=0.30\textwidth]{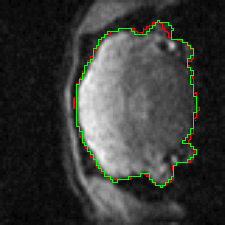}}\\
\subfloat[]{\label{fig:s04}\includegraphics[width=0.30\textwidth]{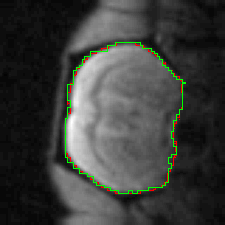}}\hfill
\subfloat[]{\label{fig:s05}\includegraphics[width=0.30\textwidth]{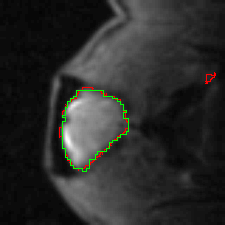}}\hfill
\subfloat[]{\label{fig:s06}\includegraphics[width=0.30\textwidth]{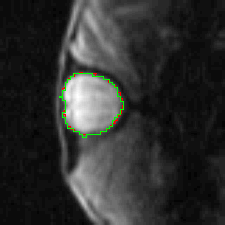}}\\
\caption{Validation results. Green line represents the ground truth and red line is the predicted region.}
\label{fig:skulls}
\end{figure*}

The same model that obtained the best BCE loss on validation dataset has 6 out of its 49 results depicted in Fig.~\ref{fig:skulls}. In general, the model performs well to segment the rodent's brain in an fMRI. There is one validation result which has a small mistake in the segmentation. That is depicted in Fig~\ref{fig:s05} and it has a small predicted region on the right side of the image which means that the model predicted a ``second" tiny brain. Despite that, the qualitative and qualitative results are impressive.

\section{Discussion}

Much of what is known in neuroscience is derived from studies using rodent models, due to its versatility and the large selection of methods (e.g., invasive methods) available to study them. On the other hand, much of what is known about the human brain is derived from MRI and fMRI studies. Thus, fMRI holds the promise of bridging the gap between what we know about the mammalian brain. It may provide evidence to generalize results from rodent-derived studies using electrophysiology, optical, and pharmacological methods to the human model. 
In this context, it is important to create powerful tools that can increase the speed and the reliability of the analysis performed on data derived from rodent models and can equally be applied to human data.
In this article we made a step towards democratizing deep learning tools to the neuroscience community by successfully applying a U-Net to perform skull stripping of low resolution functional magnetic resonance images from rodents. The method was quick to train, required little data due to the usage of data augmentation techniques, and qualitatively performed reasonably well. 
In contrast to other approaches that depend on images with high-resolution images or deformations of initial masks, U-Nets work well with low resolution images and can segment distorted images, even with motion artifacts. Additionally, by using a network that operates on images as inputs instead of a 3D tensor with all the image slices at once, we could use data augmentation strategies without major problems with respect to alignment issues.
However, we recognize that U-Net may not be the best nor the fastest architecture to perform semantic segmentation. Other topologies such as Albunet, or Ternausnet \cite{wang2018deep} might deploy better segmentation at higher speeds. Additionally, because fMRI has a temporal component, recursive layers could be added to take the dynamic nature of the signal in the brain as a feature to better segment and remove the skull, perhaps even in a non-supervised manner. Consequently, a logical step is to explore how more modern architectures could perform in this task.
We hope this tool helps neuroscientists to reduce time in preprocessing steps of their analysis of fMRI data in non-human models.

\section*{Acknowledgment}
The two authors, Annelene Gulden Dahl and Sidney Pontes-Filho contributed equally to the work. We also thank Stefano Nichele for the comments on the manuscript.
This work was supported by Norwegian Research Council SOCRATES project (grant number 270961) and received internal support as a lighthouse project in Computer Vision from the Faculty of Technology, Art and Design (TKD) at Oslo Metropolitan University, Norway.

\section*{Repository}
Code and example data can be found in the following repository:
\url{https://github.com/sidneyp/skull-stripper}

\bibliographystyle{IEEEtran}
\bibliography{IEEEabrv,bibliography}

\end{document}